\documentstyle[12pt,preprint,aps,epsfig]{revtex}
\tighten
\newcommand{\beq}{\begin{eqnarray}}
\newcommand{\eeq}{\end{eqnarray}}
\newcommand{\bsga}{ b \to s \gamma }

\newcommand{\as}{ \alpha_s }
\newcommand{\aem}{ \alpha_{em} }

\newcommand{\mw}{ M_W }

\newcommand{\mhp}{ M_{H^{+}}}
\def\etap{\eta^{\prime}}
\def\etapp{\eta^{(')}}

\newcommand{\tab}[1]{Table \ref{#1}}
\newcommand{\fig}[1]{Fig.\ref{#1}}

\newcommand{\non}{\nonumber\\ }
\title{\bf{ $\etap K $ puzzle of $B $ meson decays and new physics
effects in the general two-Higgs-doublet model } }
\author{ Zhenjun Xiao$^{(1,2)}$ \thanks{E-mail: zxiao@ibm320h.phy.pku.edu.cn},
Chong Sheng Li$^{(1)}$ \\
{\small 1. Department of Physics, Peking University, Beijing,
 100871, People's Republic of China }\\
{\small 2. Department of Physics, Henan Normal University, Xinxiang, Henan,
453002, People's Republic of China}  \\
 Kuang-Ta Chao \\
{\small  CCAST(World Laboratory), P.O.Box 8730, Beijing 100080,
People's Republic of China }\\
{\small and Department of Physics, Peking University, Beijing, 100871,
People's Republic of China}    }
\date{\today}
\begin{document}
\maketitle
\begin{abstract}
we calculate the new physics contributions to seven measured decays
$B \to \pi^+ \pi^-, K \pi$ and $ K \etap$ in the general two-Higgs-doublet
model (Model III). Within the considered parameter space, we find that: (a)
the CLEO/BaBar measurement of $B\to \pi^+ \pi^-$ decay prefers a
small $F_0^{B\pi}(0)$: $F_0^{B\pi}(0)=0.25 \pm 0.03$;
(b) the new physics enhancements  to the penguin-dominated $B \to K\pi$ and
$B\to K \etap$ decays are significant in size, $\sim (40-70)\%$
$w.r.t$ the standard model predictions;
and (c) the new physics enhancements can boost the branching ratios
${\cal B}(B \to K^+ \etap)$ and ${\cal B}(B \to K^0 \etap)$
to be consistent with the data within one standard deviation, and hence lead to
a simple and plausible new physics interpretation for the $\etap K$ puzzle.
\end{abstract}


\vspace{0.5cm} \noindent
PACS numbers: 13.25.Hw, 12.15.Ji, 12.38.Bx, 12.60.Fr

\newpage

One of the main objectives of B experiments is to probe for possible effects of new
physics beyond the standard model (SM).
Precision measurements of B meson system can provide an insight into very
high energy scales via the indirect loop effects of new physics.
The B system therefore offers a complementary probe to the search for new
physics at other hadron colliders \cite{slac504,xiao20}.

Up to now,  CLEO, BaBar and Belle Collaboration
\cite{cleo9912,cleo2000,babar2000,belle2000} have observed fourteen
$B_{u,d}$ meson two-body charmless hadronic decay modes
\beq
B\to \pi^\pm \pi^\mp,\;  K \pi,\;  K \etap,\;  \rho^\pm \pi^\mp,\;
\rho^0\pi^\pm,\;  \omega\pi^\pm,\;  K^*\eta,\;  K^{* \pm} \pi^\mp,\;  \phi K^\pm\;.
\eeq
The experimental measures are generally consistent with the theoretical predictions
based on the effective Hamiltonian with generalized factorization\cite{ali9804,chen99},
with an exception of the so-called $\etap K$ puzzle: the $B \to K\etap$ decay rates
are much larger than that expected in the SM \cite{cleo9912}.

The unexpectedly large branching ratios of $B \to K \etap $ was
firstly reported in 1997 by CLEO \cite{cleo98}, and confirmed
very recently by  CLEO and BaBar \cite{cleo9912,babar2000}:
${\cal B }(B\to K^+ \etap)=( 75 \pm 10) \times 10^{-6}$ ( average
of the CLEO and BaBar result), and ${\cal B}(B \to K^0 \etap )=
( 89 ^{+18}_{-16} \pm 9 )\times 10^{-6}$ ( CLEO ). The $K \etap$ signal
is large and  stable, and clearly much larger than the SM predictions
${\cal B}( B \to K \etap) =(20-50)\times 10^{-6}$ as given in
Refs.\cite{ali9804,cheng00a,x10326}.
In order to accommodate the data, one may need
an additional contribution unique to the $\etap $ meson in the framework of the SM,
or enhancements from new physics models beyond the SM to explain the
$B \to K \etap$ puzzle \cite{hz98}.

In a previous paper\cite{x10326}, we considered the second possibility and calculated
the new physics effects on the branching ratios of seventy six $B \to h_1 h_2$ decay
modes in the general two-Higgs-doublet models (2HDM's)
\cite{hou92,atwood97}, and found that the new physics enhancement to the
penguin-dominated decay modes can be significant.
In this letter, we focus on seven $B \to P P$ decays (where P refers to the light
pseudo-scalar mesons) whose branching ratios have already been measured.
We firstly find the constraint on the form factor $F_0^{B\pi}(0)$
from the measured $B \to \pi^\pm \pi^\mp$ decay rate, and then check the consistency
between the theoretical predictions and the data for the four $B \to K \pi $ decay
modes after including the new physics contributions in model III: the third type
of two-Higgs-doublet models\cite{hou92,atwood97}. We finally calculate the new
physics enhancements to the $B \to K \etap$ decays and study the effects of
major uncertainties.

On the theory side, one usually uses the low-energy effective Hamiltonian with generalized
factorization \cite{bbl96,bsw87,ali9804,chen99} to calculate the two-body
charmless $B$ meson decays.  For the  inclusive three-body decays $b \to s \bar{q} q $
with $q\in \{u,d,s \}$ the effective Hamiltonian can be written as \cite{ali9804},
\begin{equation}
{\cal H}_{eff}(\Delta B=1) = \frac{G_F}{\sqrt{2}} \left \{
\sum_{j=1}^2 C_j \left ( V_{ub}V_{us}^* Q_j^u  + V_{cb}V_{cs}^*
Q_j^c \right ) - V_{tb}V_{ts}^* \left [ \sum_{j=3}^{10}  C_j Q_j +
C_{g} Q_{g} \right ] \right \}~, \label{heff2}
 \end{equation}
where the operator basis contains the current-current operators $Q_{1,2}$,
the QCD penguin operators $Q_{3-6}$, the electroweak penguin operators
$Q_{9-10}$ and the chromo-magnetic dipole operator $Q_g$, the explicit
expressions can be found easily for example in Ref.\cite{ali9804}.
For $b \to d \bar{q} q$ decays, one simply makes the replacement $s \to d$.

Following Ref.\cite{ali9804}, we also neglect the effects
of the electromagnetic penguin operator $Q_{7\gamma}$, and do not
consider the effect of the weak annihilation and exchange diagrams.
The coefficients $C_{i}$ in Eq.(\ref{heff2}) are the
well-known Wilson coefficients. Within the SM and at scale $M_W$,
the Wilson coefficients $C_1(M_W), \cdots, C_{10}(M_W)$ at next-to-leading logarithmic
order {NLO) and $C_{g}(M_W)$ at leading logarithmic order (LO) have been given
for example in Ref.\cite{bbl96}.

In a recent paper \cite{chao99}, Chao {\it et al.} studied the decay $\bsga$ in
model III by assuming that only the couplings $\lambda_{tt}=|\lambda_{tt}|
e^{i\theta_t}$ and $\lambda_{bb}=|\lambda_{bb}|e^{i\theta_b}$ are
non-zero\footnote{For more details about the structure of  model III and the
available experimental constraints, one can see Refs.\cite{atwood97,xiao20,chao99}.}.
They found that the constraint on $\mhp$ imposed by the CLEO
data of $\bsga$ can be greatly relaxed by considering the phase
effects of $\lambda_{tt}$ and $\lambda_{bb}$. From the studies of
Refs.\cite{xiao20,chao99}, we know that for model III the parameter space
\beq
&& \lambda_{ij}=0, \ \ for \ \ ij\neq tt,\ \ or \ \  bb, \nonumber\\
&& |\lambda_{tt}|= 0.3,\ \ |\lambda_{bb}|=35,\ \
\theta=(0^0 - 30^0),\ \ \mhp=(200 \pm 100 ){\rm GeV}, \label{eq:lm3}
\eeq
are allowed by the available data, where $\theta=\theta_{bb}-\theta_{tt}$.
In this letter, we calculate the new physics contributions to seven B meson decay
modes in the Chao-Cheung-Keung (CCK) scenario of model III \cite{chao99}. Since the
new physics corrections on the branching ratios of two-body charmless  hadronic $B_{u,d}$
decays in models I and II are small in magnitude \cite{x10326}, we do not
consider the cases of models I and II in this letter.

Following the same procedure as in the SM, it is straightforward to calculate the new
$\gamma$-, $Z^0$- and gluonic penguin diagrams induced by the exchanges of charged-Higgs
bosons appeared  in model III \cite{x10326}.
After taking into account the new physics (NP) contributions,
the Wilson coefficients $C_i(\mw)$ $i=1,\cdots, 10$ at the NLO level
and $C_g$ at the LO level can be written as
\beq
C_1(\mw) &=& 1 - \frac{11}{6} \; \frac{\as(\mw)}{4\pi}
               - \frac{35}{18} \; \frac{\aem}{4\pi} \, , \label{eq:c1mw}\\
C_2(\mw) &=&     \frac{11}{2} \; \frac{\as(\mw)}{4\pi} \, , \\
C_3(\mw) &=& -\frac{\as(\mw)}{24\pi} \left [ E_0(x_t) +E_0^{NP} -\frac{2}{3}
\right ]\non
&& +\frac{\aem}{6\pi} \frac{1}{\sin^2\theta_W}
             \left[ 2 B_0(x_t) + C_0(x_t) + C_0^{NP} \right] \, ,\\
C_4(\mw) &=& \frac{\as(\mw)}{8\pi} \left [ E_0(x_t)+E_0^{NP}
-\frac{2}{3} \right ] \, ,\\
C_5(\mw) &=& -\frac{\as(\mw)}{24\pi}
\left [E_0(x_t)+E_0^{NP} -\frac{2}{3} \right ] \, , \\
C_6(\mw) &=& \frac{\as(\mw)}{8\pi}
\left [E_0(x_t)+E_0^{NP} -\frac{2}{3} \right ] \, ,\\
C_7(\mw) &=& \frac{\aem}{6\pi} \left [ 4 C_0(x_t) + 4 C_0^{NP}
    + D_0(x_t) +D_0^{NP} -\frac{4}{9}\right ]\, , \\
C_8(\mw) &=& C_{10}(\mw)= 0 \, , \\
C_9(\mw) &=& \frac{\aem}{6\pi} \left\{ 4C_0(x_t) + 4 C_0^{NP}
    +D_0(x_t) +D_0^{NP} -\frac{4}{9}\right. \non
&& \left. +  \frac{1}{\sin^2\theta_W} \left [ 10 B_0(x_t)
- 4 C_0(x_t) + 4 C_0^{NP} \right ]  \right\} \, ,\\
C_{g}(\mw) &=& -\frac{1}{2}\left ( E'_0(x_t) + {E'}_0^{NP} \right )
\, ,\label{eq:cimw}
\eeq
where $x_t=m_t^2/M_W^2$, the functions $B_0(x)$, $C_0(x)$, $D_0(x)$, $E_0(x)$
and $E'_0$ are the familiar Inami-Lim functions which describe
the contributions from the $W$-penguin and Box diagrams in the SM,
and can be found easily, for example, in Ref.\cite{bbl96}. The functions
$C_0^{NP}$, $D_0^{NP}$, $E_0^{NP}$ and ${E'}_0^{NP}$ in
Eq.(\ref{eq:cimw}) describe the new physics contributions to Wilson coefficients
in model III\cite{x10326},
\beq
C_0^{NP} &=& \frac{-x_t}{16} \left[ \frac{y_t}{1-y_t} + \frac{y_t}{(1-y_t)^2}\ln[y_t]
\right ]\cdot |\lambda_{tt}|^2~, \label{eq:c0m3}\\
D_0^{NP} &=& -\frac{1}{3} H(y_t)|\lambda_{tt}|^2~, \label{eq:d0m3}\\
E_0^{NP} &=& -\frac{1}{2}I(y_t)|\lambda_{tt}|^2~,\label{eq:e0m3}\\
{E'_0}^{NP}&=& \frac{1}{6}J(y_t)|\lambda_{tt}|^2
 - K(y_t) |\lambda_{tt} \lambda_{bb}| e^{i\theta}~,\label{eq:e0pm3}
\eeq
with
\beq
H(y)&=& \frac{38 y - 79 y^2 + 47 y^3}{72(1-y)^3}
+ \frac{4y -6y^2 + 3y^4}{12 (1-y)^4} \ln[y]~,\label{eq:hhy}\\
I(y)&=& \frac{16y -29y^2 +7 y^3}{36(1-y)^3} + \frac{2y- 3y^2}{6(1-y)^4}\log[y]~,
\label{eq:iiy} \\
J(y)&=& \frac{2y + 5y^2 - y^3}{4(1-y)^3} + \frac{3y^2}{2(1-y)^4}\log[y]~,
\label{eq:jjy} \\
K(y)&=& \frac{-3y + y^2}{4(1-y)^2} - \frac{y}{2(1-y)^3}\log[y]~,
\label{eq:kky}
\eeq
where $x_t=m_t^2/\mw^2$, $y_t=m_t^2/\mhp^2$, and the small terms
proportional to $m_b^2/m_t^2$ have been neglected.

Since the heavy charged Higgs bosons appeared in model III have
been integrated out at the scale $M_W$, the QCD running of the
Wilson coefficients $C_i(M_W)$ down to the scale $\mu=O(m_b)$
after including the NP contributions will be the same as in the SM.
In the NDR scheme, by using the input parameters as given in
Eqs.(\ref{eq:lm3}) and (\ref{eq:masses}) and
setting $\mu=2.5$ GeV,  we find that:
\begin{eqnarray}
&& C_1 =  1.1245,\ \   C_2 = - 0.2662, \ \ C_3  =  0.0186,\ \   C_4 = -0.0458,\non
&& C_5 =  0.0113, \ \ C_6  =  -0.0587,\ \  C_7  =  0.0006, \ \  C_8  = 0.0007,\non
&& C_9 =  -0.0096,\ \ C_{10} = 0.0026,  C_g^{eff}  =  0.3364 \label{eq:cgmb2}
\end{eqnarray}
where $C_g^{eff} = C_{8G}+ C_5$.

In this letter, the generalized factorization ansatz as being used in
Ref.\cite{chen99} will be employed. For the studied seven $B$ meson decay
modes, we use the decay amplitudes as given in Ref.\cite{ali9804} without
further discussion about details. We focus on estimating the new physics effects on
these seven measured decay modes.
In the NDR scheme and for $SU(3)_C$, the effective Wilson coefficients
can be written as \cite{chen99}
\beq
C_i^{eff} &=& \left [ 1 + \frac{\alpha_s}{4\pi} \, \left( \hat{r}_V^T +
 \gamma_{V}^T \log \frac{m_b}{\mu}\right) \right ]_{ij} \, C_j
 +\frac{\alpha_s}{24\pi} \, A_i' \left (C_t + C_p + C_g \right)
+ \frac{\alpha_{ew}}{8\pi}\, B_i' C_e ~, \label{eq:wceff}
\eeq
where $A_i'=(0,0,-1,3,-1,3,0,0,0,0)^T$, $B_i'=(0,0,0,0,0,0,1,0,1,0)^T$, the
matrices  $\hat{r}_V$ and $\gamma_V$ contain the process-independent
contributions from the vertex diagrams. The matrix $\gamma_V$
and $\hat{r}_V$ have been given explicitly, for example, in Eq.(2.17) and (2.18) of
Ref.\cite{chen99} \footnote{The correct value of the element $(\hat{r}_{NDR})_{66}$
and $(\hat{r}_{NDR})_{88}$ should be  17 instead of 1 as pointed in
Ref.\cite{cheng00a}.}.
The function $C_t$, $C_p$, and $C_g$ describe the contributions arising
from the penguin diagrams of the current-current
$Q_{1,2}$, the QCD operators $Q_3$-$Q_6$, and the tree-level diagram of the
magnetic dipole operator $Q_{8G}$, respectively. The explicit expressions of
the functions $C_t$, $C_p$, and $C_g$ can be found for example in
Refs.\cite{chen99,x10326}.
We here  also follow the procedure of Ref.\cite{ali98} to include the contribution
of magnetic gluon penguin.

In the generalized factorization ansatz, the effective Wilson coefficients $C_i^{eff}$
will appear in the decay amplitudes in the combinations,
\begin{equation}
a_{2i-1}\equiv C_{2i-1}^{eff} +\frac{{C}_{2i}^{eff}}{N_c^{eff}}, \ \
a_{2i}\equiv C_{2i}^{eff}     +\frac{{C}_{2i-1}^{eff}}{N_c^{eff}}, \ \ \
(i=1,\ldots,5) \label{eq:ai}
\end{equation}
where the effective number of colors $N_c^{eff}$ is treated as a free parameter
varying in the range of $2 \leq N_c^{eff} \leq \infty$, in order to
model the non-factorizable contribution to the hadronic matrix elements.
It is evident that  the reliability of generalized
factorization approach has been improved since the effective Wilson
coefficients $C_i^{eff}$ appeared in Eq.(\ref{eq:ai}) are now gauge
invariant and infrared safe\cite{cheng99a}. Although  $N_c^{eff}$ can in principle
vary from channel to channel, but in the energetic two-body hadronic B
meson decays, it is expected to be process insensitive as supported by
the data \cite{chen99}.

In the B rest frame, the branching ratios ${\cal B}(B \to PP)$ can
be written as
\begin{equation}
{\cal B}(B \to X Y )=  \frac{1}{\Gamma_{tot}} \frac{|p|}{8\pi M_B^2}
|M(B\to XY)|^2~,\label{eq:brbpp}
\end{equation}
where $\Gamma_{tot}(B_u^-)=3.982 \times 10^{-13}$ GeV
and $\Gamma_{tot}(B_d^0)=4.252 \times 10^{-13}$GeV obtained by using
$\tau(B_u^-)=1.653 ps$ and $\tau(B_d^0)=1.548 ps$ \cite{pdg2000}, $p_B$ is the
four-momentum of the B meson, $M_B=5.279$ GeV is the mass of $B_u$ or $B_d$ meson,
and
\begin{equation}
|p| =\frac{1}{2M_B}\sqrt{[M_B^2 -(M_X + M_Y)^2] [ M_B^2 -(M_X-M_Y)^2 ]}
\label{eq:pxy}
\end{equation}
is the magnitude of momentum of particle X and Y in the B rest frame.

In the numerical calculations the following input parameters will be used:

\begin{itemize}
\item
The coupling constants, gauge boson masses, light meson masses, $\cdots$,
(all masses in unit of GeV )\cite{ali9804,pdg2000}
\beq
&& \alpha_{em}=1/128, \;  \alpha_s(M_Z)=0.118,\; \sin^2\theta_W=0.23,\; G_F=1.16639\times 10^{-5} (GeV)^{-2}, \non
&& M_Z=91.188, \;   M_W=80.42,\; m_{B_d^0}=m_{B_u^\pm}=5.279,\; m_{\pi^\pm}=0.140,\;\non
&& m_{\pi^0}=0.135,\;   m_{\eta}=0.547,\; m_{\etap}=0.958,\; m_{K^\pm}=0.494,\;  m_{K^0}=0.498.
\label{eq:masses}
\eeq

\item
The elements of CKM matrix in the Wolfenstein parametrization: $A=0.81$,
$\lambda=0.2205$, $\rho=0.12$ and $\eta=0.34$ (which corresponds to $\gamma=71^\circ$
and $\beta=26^\circ$), and the uncertainty of $\delta \eta =\pm 0.08$ will be
considered.

\item
We firstly treat the internal quark masses in the loops as constituent masses,
\beq
m_b=4.88 GeV, \; m_c=1.5 GeV,\; m_s=0.5 GeV, \; m_u=m_d=0.2
GeV. \label{eq:massa}
\eeq
Secondly, we use the current quark
masses for $m_i$ ($i=u,d,s,c,b$) which appear through the equation
of motion when working out the hadronic matrix elements. For
$\mu=2.5 GeV$, one finds\cite{ali9804}
\beq m_b=4.88 GeV, \;
m_c=1.5 GeV, m_s=0.122 GeV, \; m_d=7.6 MeV,\; m_u=4.2 MeV. \label{eq:massb}
\eeq
For the mass of heavy top quark we also use $m_t=\overline{m_t}(m_t)=168 GeV$.

\item
The decay constants of light mesons (in the units of MeV) are
\beq
&&f_{\pi}=133,\; f_{K}=158,\; f^u_{\eta}=f^d_{\eta}=78,\;
f^u_{\etap}=f^d_{\etap}=68,\non
&&f^c_{\eta}=-0.9,\; f^c_{\etap}=-0.23, f^s_{\eta}=-113,\; f^c_{\etap}=141. \label{fpis}
\eeq
where $f^u_{\etapp}$ and $f^s_{\etapp}$ have been defined in the two-angle-mixing
formalism with $\theta_0=-9.1^\circ$ and $\theta_8 =-22.2^\circ$\cite{fk97}.

\item
The form factors at the zero momentum transfer  are
\beq
F_0^{B\pi}(0)=0.33,\; F_0^{BK}(0)=0.38,\; F_0^{B\eta}(0)=0.145,\;
F_0^{B\etap}(0)=0.135\label{eq:bsw}
\eeq
in the BSW model \cite{bsw87}, and
\beq
F_0^{B\pi}(0)=0.36,\; F_0^{BK}(0)=0.41,\; F_0^{B\eta}(0)=0.16,\;
F_0^{B\etap}(0)=0.145,
\label{eq:lqqsr}
\eeq
in the LQQSR approach \cite{ali9804}. Here the relation between $F_0^{B\etap}(0)$ and
$F_0^{B\pi}(0)$ as defined in Eq.(A12) in Ref.\cite{ali9804} has been used.
The momentum dependence of $F_0(k^2)$ as defined in Ref.\cite{bsw87} is
$ F_0(k^2)= F_0(0)/(1-k^2/m^2(0^+))$. The pole masses being used to evaluate the $k^2$-dependence
of form factors are $m(0^+)=5.73$ GeV for $\bar{u}b$ and $ \bar{d}b$ currents, and
$ m(0^+)= 5.89$ GeV for $\bar{s}b $ currents.

\end{itemize}

For the seven studied $B$ meson decay modes, currently available measurements from
CLEO, BaBar and Belle Collaboration \cite{cleo9912,cleo2000,babar2000,belle2000}
are as follows:
\beq
{\cal B}(B \to \pi^+ \pi^-)&=& \left \{\begin{array}{ll}
( 4.3 ^{+1.6}_{-1.5} \pm 0.5 )\times 10^{-6} & {\rm [CLEO]}, \\
( 9.3 ^{+2.8\; +1.2 }_{-2.1\; -1.4} )\times 10^{-6} & {\rm [BaBar]},  \\
\end{array} \right. \label{eq:brexp01} \\
{\cal B}(B \to K^+ \pi^0)&=& \left \{\begin{array}{ll}
( 11.6 ^{+3.0\; +1.4}_{-2.7\; -1.3} )\times 10^{-6} & {\rm [CLEO]}, \\
( 18.8 ^{+5.5}_{-4.9} \pm 2.3 )\times 10^{-6} & {\rm [Belle]},  \\
\end{array} \right. \label{eq:brexp11} \\
{\cal B}(B \to K^+ \pi^-)&=& \left \{\begin{array}{ll}
( 17.2 ^{+2.5}_{-2.4} \pm 1.2)\times 10^{-6} & {\rm [CLEO]}, \\
( 12.5 ^{+3.0\; +1.3}_{-2.6\; -1.7} \pm 2.3 )\times 10^{-6} & {\rm [BaBar]},  \\
( 17.4 ^{+5.1}_{-4.6} \pm 3.4)\times 10^{-6} & {\rm [BELLE]}, \\
\end{array} \right. \label{eq:brexp12} \\
{\cal B}(B \to K^0 \pi^+)&=&
( 18.2 ^{+4.6}_{-4.0} \pm 1.6)\times 10^{-6}\ \   {\rm [CLEO]},
\label{eq:brexp13}\\
{\cal B}(B \to K^0 \pi^0)&=& \left \{\begin{array}{ll}
( 14.6 ^{+5.9\; +2.4}_{-5.1\; -3.3} )\times 10^{-6} & {\rm [CLEO]}, \\
( 21  ^{+9.3\; +2.5}_{-7.8\; -2.3} )\times 10^{-6} & {\rm [BELLE]},  \\
\end{array} \right. \label{eq:brexp14} \\
{\cal B}(B \to K^+ \etap )&=& \left \{\begin{array}{ll}
( 80 ^{+10}_{-9} \pm 7 )\times 10^{-6} & {\rm [CLEO]}, \\
( 62 \pm 18 \pm 8 )\times 10^{-6} & {\rm [BaBar]},  \\
\end{array} \right. \label{eq:brexp16} \\
{\cal B}(B \to K^0 \etap )&=&
( 89 ^{+18}_{-16} \pm 9 )\times 10^{-6} \ \  {\rm [CLEO]}.
\label{eq:brexp18}
\eeq
The measurements of CLEO, BaBar and BELLE Collaboration are consistent with each
other within errors.

In Table \ref{bpp1}, we present the theoretical predictions of
the branching ratios for the seven $B$ decay modes in the framework
of the SM and model III by using the form factors from Bauer, Stech
and Wirbel (BSW) model  \cite{bsw87} and Lattice QCD/QCD sum rule (LQQSR)
model  \cite{flynn97}, as listed in the first  and second entries respectively.
The last column shows the CLEO data of $B\to K^0 \pi^+, K^0\etap$ decays,
and the average of CLEO, BaBar and/or BELLE measurements for other five
decay modes. Theoretical predictions are made by using the central values of
input parameters as given in Eqs.(\ref{eq:lm3},\ref{eq:masses}-\ref{eq:lqqsr}),
and assuming $\mhp=200$GeV and $N_c^{eff}=2, 3,
\infty$ in the generalized factorization approach.
The branching ratios collected in \tab{bpp1} are the averages of
the branching ratios of $B$ and anti-$B$ decays. The ratio $\delta
{\cal  B}$  describes the new physics correction on the decay
ratio and is defined as
\begin{equation}
\delta {\cal  B} (B \to XY) = \frac{{\cal  B}(B \to XY)^{III}
-{\cal  B}(B \to XY)^{SM}}{{\cal  B}(B \to XY)^{SM}} \label{eq:dbr}
\end{equation}
The last column in shows the CLEO data of $B\to K^0 \pi^+, K^0\etap$ decays,
and the average of CLEO, BaBar and/or BELLE measurements for remaining four
decay modes.

From \tab{bpp1}, we find that
\begin{itemize}
\item
The SM prediction of $B^0 \to \pi^+\pi^-$ decay is clearly much larger than
the CLEO measurement, but agree with BaBar measurement. The new physics
contribution to this tree-dominated decay mode, however,  is negligibly small.

\item
For four $B \to K \pi$ decays, the SM predictions agree with experimental
measurements within errors. In model III, the new physics enhancements are large in magnitude:
$\sim 50\%$ $w.r.t$ the SM predictions. The model III predictions are generally
larger than the data for $B \to K^+ \pi, K^0\pi^+$ decays but still agree with
the data within $2\sigma$ errors since both the theoretical and experimental
errors are still large now.

\item
For $B \to K\etap$ decays, the new physics enhancements in model III are large in
magnitude: $\sim 60\%$ $w.r.t$ the SM predictions. Such enhancement can make the
theoretical predictions become consistent with the CLEO/BaBar data within one
standard deviation, as illustrated in \fig{fig:fig1} where the dot-dashed
and solid curve shows the theoretical prediction in the model III for
$N_c^{eff}=3,\infty$, respectively.

\item
Since the form factors in LQQSR approach are larger than those in the BSW model,
the theoretical predictions in the LQQSR approach are generally larger  than
those in the BSW model by $\sim 15\%$.

\end{itemize}

Because the branching ratios of the studied B decay modes strongly depend on
the values of involved form factors $F_0^{B\pi}(0)$, $F_0^{BK}(0)$ and
$F_0^{B\etap}(0)$, any information about these form factors from data will help us
to refine the theoretical predictions.
Since the $B^0 \to \pi^+\pi^-$ decay is a tree-dominated decay mode and the possible
new physics effect is also negligibly small,  the experimental measures of this decay
lead to a stringent constraint on the form factor $F_0^{B\pi}(0)$.
If we take the average of CLEO and BaBar measurements,
\beq
{\cal B}(B_d^0\to \pi^+ \pi^-)=(5.5 \pm 1.5) \times 10^{-6},
\label{eq:pipi}
\eeq
as the experimental result, then the constraint on $F_0^{B\pi}(0)$ will be
\beq
F_0^{B\pi}(0)=0.25 \pm 0.03
\eeq
by setting $A=0.2205$, $\lambda=0.81$, $\rho=0.12$, $\eta=0.34$, $N_c^{eff}=3$,
and by neglecting FSI also.

Theoretically, small form factor  $F_0^{B\pi}(0)$ will lead to small predicted
branching ratios of $B \to K \pi$ and $B \to K \etap$ decays.
First, the relation between $F_0^{B\etap}(0)$ and $F_0^{B\pi}(0)$
as given in Ref.\cite{ali9804} is
\beq
F_0^{B\etap}(0)=F_0^{B\pi}(0) (\sin \theta_8/\sqrt{6}
+ \cos \theta_0/\sqrt{3})\label{eq:f0beta}
\eeq
with $\theta_0=-9,1^\circ$ and $\theta_8=-22.2^\circ$.
A small $F_0^{B\pi}(0)$ leads to a small $F_0^{B\etap}$ and in turn small branching
ratios of $B \to K \etap$ decays. Second, $F_0^{BK}(0)$ cannot  deviate too much from
$F_0^{B\pi}(0)$, otherwise the $SU(3)$-symmetry relation
$F_0^{B\pi} \approx F_0^{BK}$ will be badly broken.  In
\tab{bpp2}, we show the branching ratios of seven studied decay modes
obtained by using $F_0^{B\pi}(0)=0.25$ instead of $0.33$ while keep all
other input parameters remain the same as being used in \tab{bpp1}.

Contrary to the case of using $F_0^{B\pi}(0)=0.33$ in the BSW model,
where the inclusion of new physics contributions in the model III will
degenerate the
agreement between the theoretical predictions and the data for first three
$B\to K \pi$ decay modes, the inclusion of $\sim 50\%$ new physics enhancements to
$B \to K \pi$ decays for the case of using $F_0^{B\pi}(0)=0.25$ does
improve the agreement between the theory and the data, as illustrated
in Figs.(\ref{fig:fig2},\ref{fig:fig3}) where the short-dashed and
solid curve shows the predictions in the SM
and model III for the case of using $F_0^{B\pi}(0)=0.25$.
The horizontal band between two dots lines corresponds to the (averaged ) data
with $2\sigma$ errors.

For the decay $B \to K^0 \pi^0$ we find ${\cal B}(B\to K^0\pi^0) = (4.3 \pm 2.1)
\times 10^{-6} \times ( F_0^{B\pi}(0)/0.25)^2$ in the SM, which is almost four
times smaller than the central value of the averaged data: ${\cal B}(B\to K^0 \pi^0)
=(16.6 \pm 5.3)\times 10^{-6}$. The sixty percent new physics enhancement will be
helpful to increase the theoretical prediction, but   is still
not large enough to cover the gap, as illustrated in Fig.(3b) where the lower
short-dashed and solid curves show the theoretical prediction in the SM and
model III with $F_0^{B\pi}(0)=0.25$. This problem will become clear when more
precise data from B factories are available.

For $F_0^{B\pi}(0)=0.25$, the SM predictions for branching ratios
of $B \to K \etap$ decays are in the range of $(18-40)\times 10^{-6}$ as shown in
\tab{bpp2} and clearly much smaller than the data.
We know that the $K\etap$ decay rates can be enhanced, for example,  through (i)
constructive interference in gluonic penguin diagrams, which is qualitatively
OK but numerical problems remain; (ii) the small running mass $m_s$ at the
scale $m_b$ \footnote{However, a rather small $m_s$ is not consistent with recent
lattice calculations.}; (iii) larger form factor $F_0^{B\etap}(0)$ due to the smaller
$\eta-\etap$ mixing angle $-15.4^0$ rather than $\approx -20^\circ$;
(iv) contribution from the intrinsic charm content of $\etap$ \cite{chao97}}.
However, as pointed out in Ref.\cite{ali98}, the above
mentioned enhancement is partially washed out by the anomaly effects in the
matrix element of pseudo-scalar densities, an effect overlooked before. As a
consequence, the net enhancement may be not large enough.
For a smaller $F_0^{B\pi}(0)=0.25\pm 0.03$ preferred by the data, the discrepancy between the
data and the SM predictions for $B \to K \etap$ decays becomes larger.

In the model III, however, the new gluonic and electroweak penguin diagrams
contribute to the $B \to K^+ \etap$ and $K^0\etap$ decays
through constructive interference with their SM counterparts and consequently
provide the large enhancements, $\sim 60\%$ $w.r.t.$ the SM
predictions, to make the theoretical predictions become consistent with the data
even for $F_0^{B\pi}(0)=0.25$ instead of $0.33$ as shown in \tab{bpp2} and
\fig{fig:fig4}.

In \fig{fig:fig4}, we plot the mass-dependence of ${\cal B}(B^+ \to K^+
\etap)$ and ${\cal B}(B^0 \to K^0 \etap)$ in the SM and model III by using
$F_0^{B\pi}(0)=0.25$ instead of $0.33$ (while all other input parameters are the
same as in \fig{fig:fig1}). The short-dashed line in \fig{fig:fig4} shows the
SM predictions
with $N_c^{eff}=3$. The dot-dashed and solid curve refer to the branching
ratios in the model III for $N_c^{eff}=3$ and $\infty$, respectively.
The upper dots band corresponds to the data with $2\sigma$ errors:
${\cal B}(B \to K^+\etap)=(75\pm 20)\times 10^{-6}$ and
${\cal B}(B \to K^0\etap)=(89 ^{+40}_{-36})\times 10^{-6}$.

By comparing the curves in Fig.\ref{fig:fig1} and Fig.\ref{fig:fig4}, it is easy
to see that (a) the gap between the SM predictions of $B \to K \etap$
decay rates and the data is enlarged by using $F_0^{B\pi}(0)=0.25$ instead of
$0.33$; (b) the new physics enhancement therefore becomes essential for the
theoretical predictions to be consistent with CLEO/BaBar result
within one standard deviation.

We know that the calculation of charmless hadronic B meson  decay rates suffers
from many theoretical uncertainties\cite{ali9804,chen99}.
Most of them have been considered in our calculation.
If we consider effects induced by the uncertainties of major input parameters
$\eta=0.34\pm 0.08$, $k^2=m_b^2/2 \pm 2$ GeV$^2$, $F_0^{B\pi}(0)=0.25\pm 0.03$,
$F_0^{BK}(0)=0.30\pm 0.05$, $0\leq 1/N_c^{eff} \leq 0.5$,  $\mhp=200\pm 100$ GeV
and $m_s=0.1-0.122$ GeV, we find numerically that
\beq
{\cal B}(B \to K^+ \etap)\approx {\cal B}(B \to K^0 \etap)
= \left \{\begin{array}{ll}
( 17 - 50 )\times 10^{-6} & {\rm in \ \ SM}, \\
( 28 - 75 )\times 10^{-6} & {\rm in \ \ model \ \ III}.  \\
\end{array} \right. \label{eq:ketap}
\eeq
The SM prediction of $B \to K\etap$ is much smaller than the
data, while the model III prediction can be consistent with the data within one
standard deviation. This is a simple and plausible new physics interpretation
for the observed $\etap K$ puzzle.

As simple illustrations we show explicitly the
dependence of the branching ratios ${\cal B}(B \to K \etap)$ on the form factor
$F_0^{BK}(0)$ and the running quark mass $m_s$ in Figs.(5,6).
In \fig{fig:fig5}, we plot the $F_0^{BK}(0)$ dependence of
the ratios ${\cal B}(B \to K^+ \etap)$ and ${\cal B}(B \to K^0 \etap)$
in the SM and model III, assuming
$\mhp=200$ GeV, $F_0^{B\pi}(0)=0.25$ and $F_0^{BK}(0)=0.25-0.40$.
The short-dashed line shows the SM predictions with $N_c^{eff}=3$.
The dot-dashed and solid curve refer to the branching ratios in model III
for $N_c^{eff}=3$ and $\infty$, respectively.
The dots band corresponds to the (averaged) data with $2\sigma$ errors.
The theoretical predictions also show a strong dependence upon
$F_0^{BK}(0)$: ${\cal B}(B \to K^+ \etap)=13.8\times
10^{-6}$ and $26.3 \times 10^{-6}$ in the SM for $N_c^{eff}=3$ and
$F_0^{BK}(0)=0.25$ and $0.40$, respectively.

In \fig{fig:fig6}, we plot the $m_s$-dependence of the ratios
${\cal B}(B \to K^+ \etap )$ and ${\cal B}(B \to K^0 \etap)$ in the SM and
model III with $F_0^{B\pi}(0)=0.25$, $F_0^{BK}(0)=0.33$ and $\mhp=200$ GeV.
The short-dashed line shows the SM predictions with $N_c^{eff}=3$.
The dot-dashed and solid curve refer to the branching ratios in  model III
for $N_c^{eff}=3$ and $\infty$, respectively.
The dots band corresponds to the (averaged) data with $2\sigma$ errors.
The theoretical predictions show a very strong dependence upon the mass
$m_s$: ${\cal B}(B \to K^+ \etap)=44.8\times
10^{-6}$ and $18.6 \times 10^{-6}$ in the SM for $N_c^{eff}=3$ and
$m_s=0.08$ GeV and $0.15$ GeV, respectively.

In short, we here studied the new physics contributions to the seven observed
$B \to P P$ decay modes, and made an effort to find a new physics interpretation
for the so-called $\etap K$ puzzle of B meson decays by employing the effective
Hamiltonian with generalized factorization.
Within the considered parameter space we found that:
\begin{itemize}
\item
The new physics enhancement  is negligibly small to tree-dominated $B \to \pi^+
\pi^-$ decay, but  can be significant to the penguin-dominated $B \to K\pi$ and
$B\to K \etap$ decay modes, $\sim (40-70)\%$ $w.r.t$ the SM predictions.

\item
The CLEO/BaBar measurement of $B\to \pi^+ \pi^-$ decay prefers a
small $F_0^{B\pi}(0)$: $F_0^{B\pi}(0)=0.25 \pm 0.03$ instead of
$0.33$ or $0.36$ in the BSW and LQQSR form factors. A smaller $F_0^{B\pi}(0)$
will leads to smaller predictions for other six $B \to K\pi$ and
$B \to K\etap$ decay modes studied here. The new physics
enhancements to $B \to K \pi$ decays are helpful to improve the
agreement between the data and theoretical predictions for these
decays.

\item
The new physics enhancements can boost the theoretical predictions of the branching
ratios ${\cal B}(B \to K^+ \etap)$ and ${\cal B}(B \to K^0 \etap)$ to be consistent
with the data within one standard deviation. This is a simple and plausible new
physics interpretation for the observed $\etap K$ puzzle.

\end{itemize}

\section*{ACKNOWLEDGMENTS}
C.S. Li and K.T. Chao  acknowledge the support by the National Natural
Science Foundation of China,  the State Commission of Science
and technology of China and the Doctoral Program Foundation of Institution
of Higher Education. Z.J. Xiao acknowledges the support by the National
Natural Science Foundation of China under the Grant No.19575015 and
10075013, and by the Excellent Young Teachers Program of Ministry of Education,
P.R.China.

\newpage
\begin{table}
\begin{center}
\caption{Branching ratios (in units of $10^{-6}$) of seven studied $B $ decay modes
in the SM and model III by using the BSW ( the first entries) and
LQQSR (the second entries) form factors, with $k^2=m_b^2/2$,
$A=0.81$, $\lambda=0.2205$, $\rho=0.12$, $\eta=0.34$, $\theta=0^\circ$,
$N_c^{eff}=2,\; 3,\; \infty$ and $\mhp=200$ GeV.
The last column shows the (averaged) data.}
\label{bpp1}
\vspace{0.2cm}
\begin{tabular} {l ccc ccc ccc l} \hline \hline
 &  \multicolumn{3}{c}{SM }&
\multicolumn{3}{c}{Model III}& \multicolumn{3}{c}{$\delta {\cal  B} \; [\%]$} & Data  \\
\cline{2-11}
Channel & $2$& $3$ & $\infty$ & $2$& $3$ & $\infty$&$2$&$3$& $\infty$&  \\ \hline
$B^0 \to \pi^+ \pi^-$       & $9.03$ &$10.3$&$12.9$&$9.26$ &$10.5$&$13.2$&$2.5$&$2.5$&$2.4$ & $5.5\pm 1.5$ \\
                            & $10.7$ &$12.2$&$15.4$&$11.0$ &$12.5$&$15.8$&$2.5$&$2.5$&$2.4$ &  \\
$B^+ \to K^+ \pi^0$         & $12.1$ &$13.5$&$16.7$&$17.4$ &$19.6$&$24.4$&$45 $&$45 $&$46 $ & $13.3\pm 2.9$ \\
                            & $14.3$ &$16.0$&$19.8$&$20.7$ &$23.3$&$29.0$&$45 $&$45 $&$46 $ &  \\
$B^0 \to K^+ \pi^-$         & $17.7$ &$19.6$&$23.8$&$26.6$ &$29.7$&$36.3$&$51 $&$51 $&$53 $ & $15.9\pm 2.2$ \\
                            & $21.0$ &$23.3$&$28.3$&$31.7$ &$35.3$&$43.1$&$51 $&$51 $&$53 $ &  \\
$B^+ \to K^0 \pi^+$         & $20.0$ &$23.3$&$30.7$&$29.9$ &$34.7$&$45.4$&$50 $&$49 $&$48 $ & $18.2 ^{+4.9}_{-4.3}$\\
                            & $23.8$ &$27.7$&$36.5$&$35.6$ &$41.3$&$54.0$&$50 $&$49 $&$48 $ & \\
$B^0 \to K^0 \pi^0$         & $7.22$ &$8.25$&$10.6$&$11.3$ &$12.9$&$16.5$&$57 $&$57 $&$56 $ & $16.6 \pm 5.3$ \\
                            & $8.61$ &$9.85$&$12.6$&$13.5$ &$15.4$&$19.7$&$57 $&$57 $&$56 $ & \\
$ B^+ \to  K^+ \eta^\prime$ & $22.9$ &$28.8$&$42.9$&$38.5$ &$47.5$&$68.3$&$68 $&$65 $&$59 $ & $75 \pm 10$\\
                            & $26.3$ &$33.1$&$49.3$&$42.3$ &$52.8$&$78.5$&$69 $&$65 $&$59 $ &   \\
$B^0 \to K^0 \eta^\prime$   & $22.0$ &$28.3$&$43.1$&$36.8$ &$46.0$&$67.5$&$67 $&$63 $&$57 $ & $89^{+20}_{-18}$ \\
                            & $25.3$ &$32.4$&$49.5$&$42.3$ &$52.8$&$77.6$&$67 $&$63 $&$57 $ &  \\
\hline
\end{tabular}\end{center}
\end{table}

\begin{table}
\begin{center}
\caption{Branching ratios (in units of $10^{-6}$) of seven studied $B $ decay modes
in the SM and Model III  by using the BSW form factors with $F_0^{B\pi}(0)=0.25$ instead
of $F_0^{B\pi}(0)=0.33$, and assuming $k^2=m_b^2/2$,
$A=0.81$, $\lambda=0.2205$, $\rho=0.12$, $\eta=0.34$, $N_c^{eff}=2,\; 3,\;
\infty$, $\theta=0^\circ$ and $\mhp=200$ GeV.
The last column shows the (averaged) data. }
\label{bpp2}
\vspace{0.2cm}
\begin{tabular} {l ccc ccc ccc l} \hline \hline
 &  \multicolumn{3}{c}{SM }&
\multicolumn{3}{c}{Model III}& \multicolumn{3}{c}{$\delta {\cal  B} \; [\%]$} & Data   \\
\cline{2-11}
Channel & $2$& $3$ & $\infty$ & $2$& $3$ & $\infty$&$2$&$3$& $\infty$ & \\ \hline
$B^0 \to \pi^+ \pi^-$       & $5.18$ &$5.89$&$7.42$&$5.31$ &$6.03$&$7.60$&$2.5$&$2.5$&$2.4$& $5.5\pm 1.5$ \\
$B^+ \to K^+ \pi^0$         & $7.44$ &$8.36$&$10.4$&$10.6$ &$12.0$&$15.0$&$43 $&$43 $&$44$&$13.3\pm 2.9$ \\
$B^0 \to K^+ \pi^-$         & $10.1$ &$11.2$&$13.6$&$15.3$ &$17.0$&$20.8$&$51 $&$51 $&$53$&$15.9\pm 2.2$ \\
$B^+ \to K^0 \pi^+$         & $11.5$ &$13.4$&$17.6$&$17.2$ &$19.9$&$26.0$&$50 $&$49 $&$48$&$18.2 ^{+4.9}_{-4.3}$\\
$B^0 \to K^0 \pi^0$         & $3.79$ &$4.29$&$5.46$&$6.04$ &$6.85$&$8.70$&$60 $&$60 $&$60$&$16.6 \pm 5.3$ \\
$ B^+ \to  K^+ \eta^\prime$ & $19.1$ &$24.4$&$36.9$&$32.5$ &$40.4$&$58.9$&$70 $&$66 $&$60$&$75 \pm 10$\\
$B^0 \to K^0 \eta^\prime$   & $18.3$ &$23.7$&$36.6$&$30.9$ &$38.9$&$57.6$&$69 $&$64 $&$58$&$89^{+20}_{-18}$ \\
\hline
\end{tabular}\end{center}
\end{table}

\newpage
\begin{figure}[t] 
\vspace{-60pt}
\begin{minipage}[t]{0.95\textwidth}
\centerline{\epsfxsize=\textwidth \epsffile{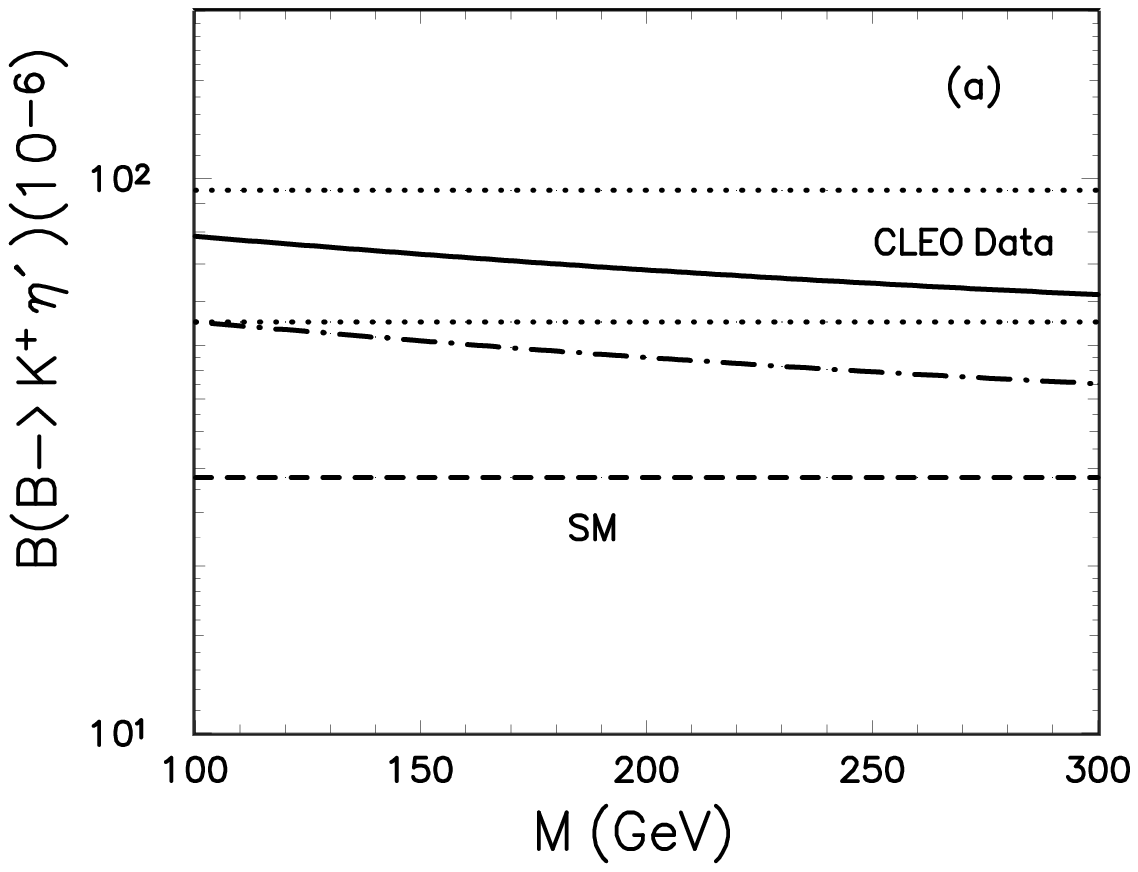}}
\vspace{-60pt}
\centerline{\epsfxsize=\textwidth \epsffile{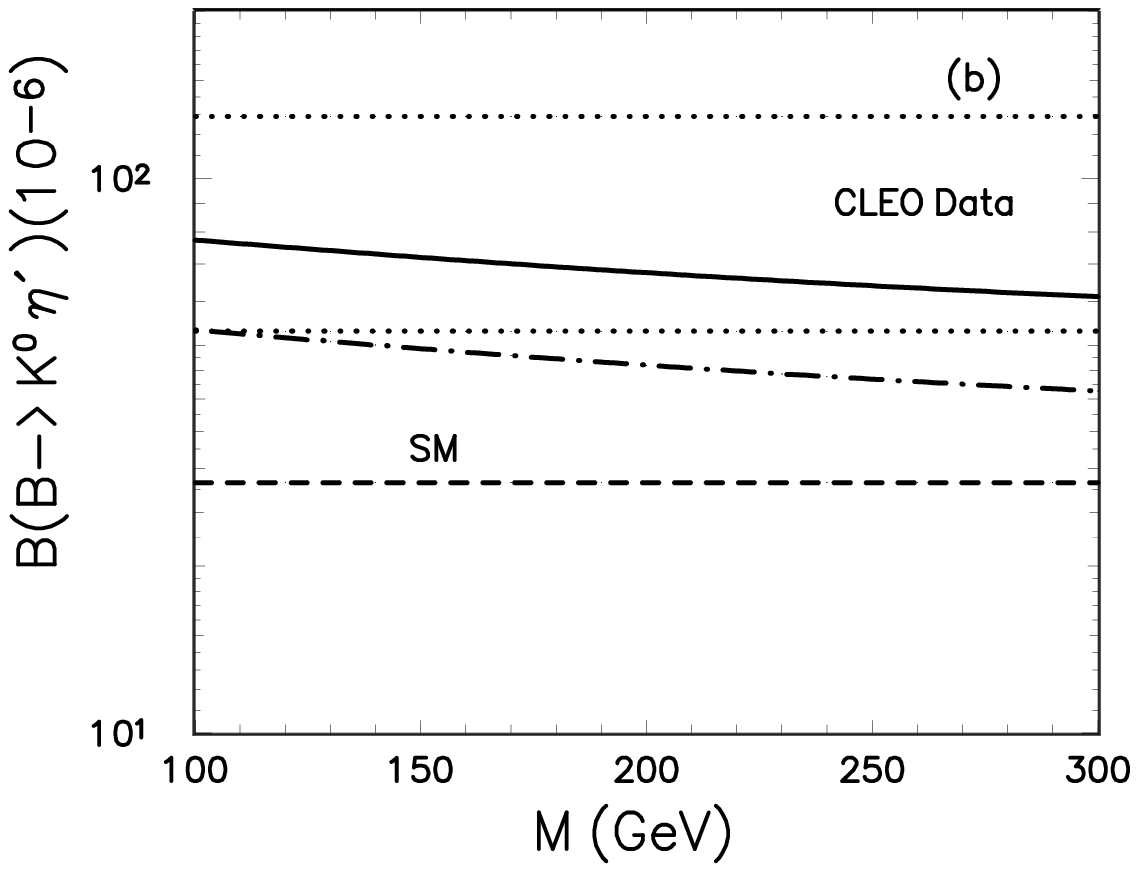}}
\caption{Plots of branching ratios of decays $B^+ \to K^+ \etap$ (1a) and
$B^0 \to K^0 \etap$ (1b) versus mass $\mhp$ in the SM and model III
with $F_0^{B\pi}(0)=0.33$.
The short-dashed line shows the SM predictions with $N_c^{eff}=3$.
The dot-dashed and solid curve refers to the branching ratios in the model III
for $N_c^{eff}=3$ and $\infty$, respectively.
The dots band corresponds to the CLEO/BaBar data with $2\sigma$ errors. }
\label{fig:fig1}
\end{minipage}
\end{figure}

\newpage
\begin{figure}[t] 
\vspace{-60pt}
\begin{minipage}[t]{0.95\textwidth}
\centerline{\epsfxsize=\textwidth \epsffile{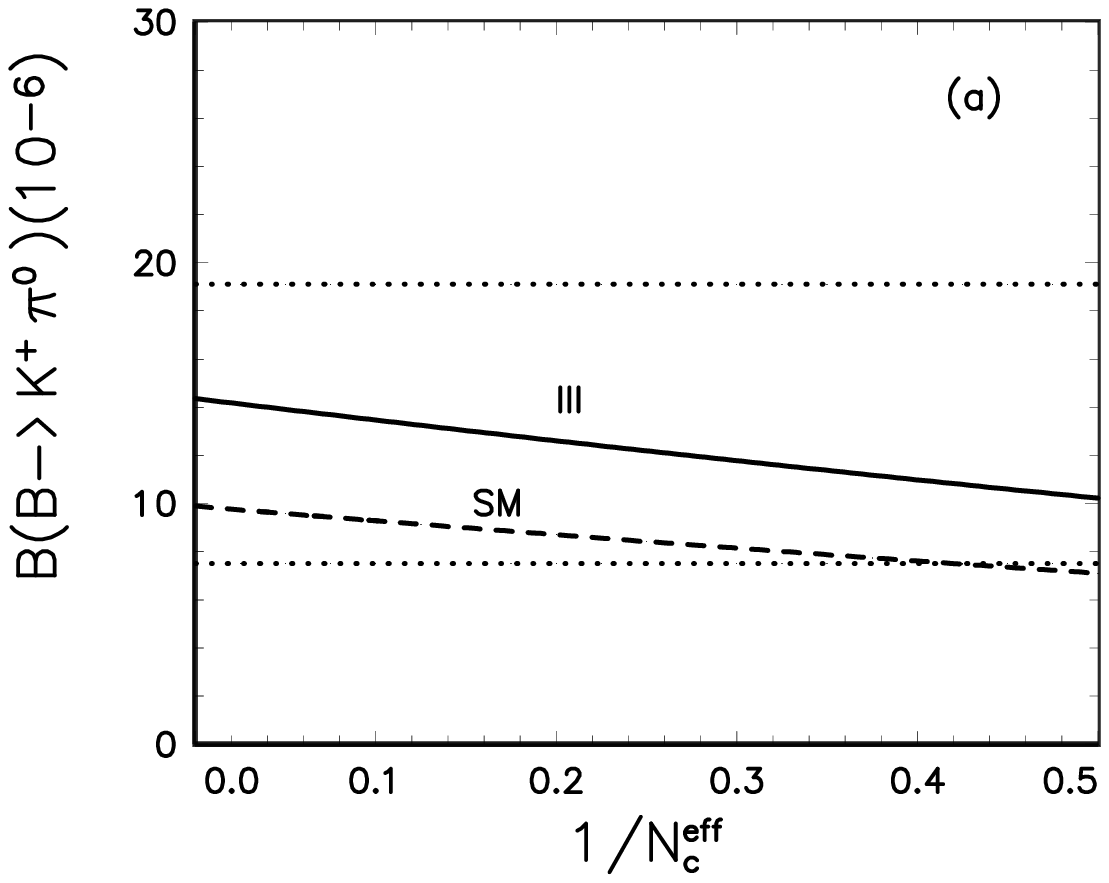}}
\vspace{-60pt}
\centerline{\epsfxsize=\textwidth \epsffile{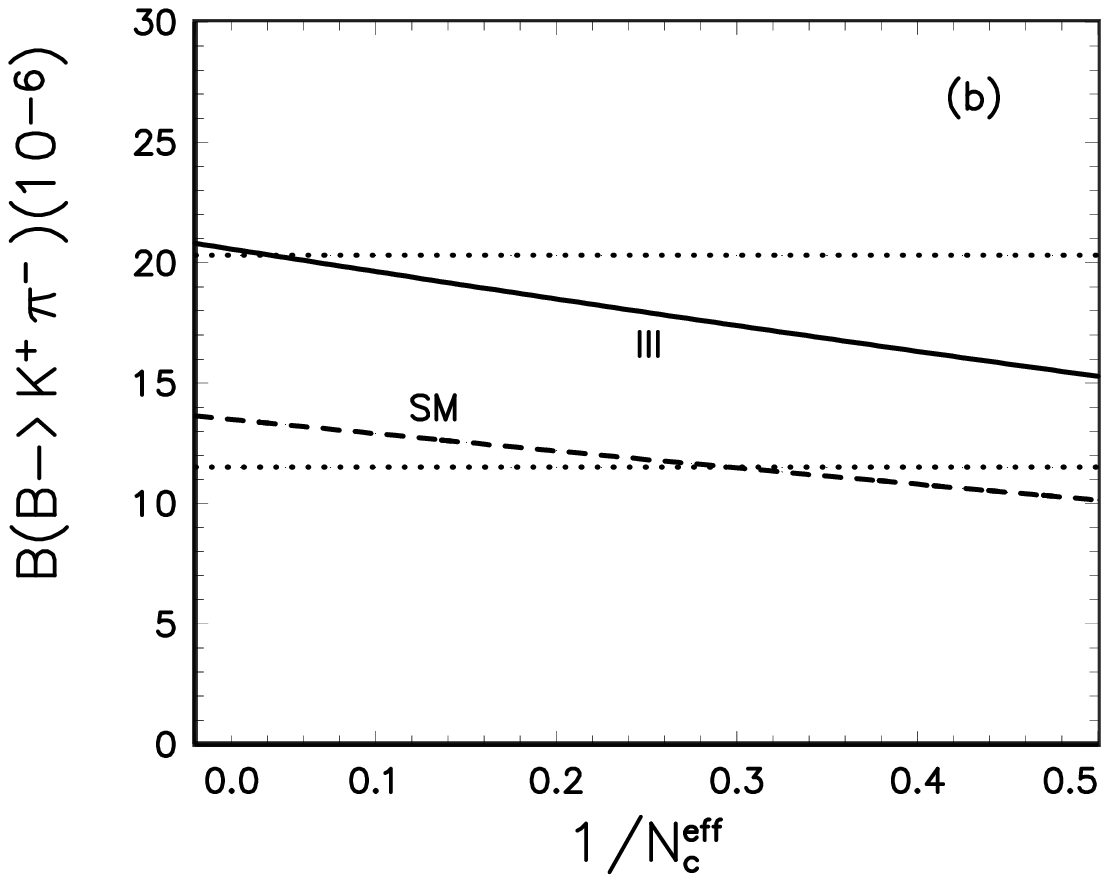}}
\caption{ Plots of branching ratios of decays $B \to K^+ \pi^0$ (2a) and
$K^+ \pi^-$ (2b) versus $1/N_c^{eff}$ in the SM and model III
assuming $\mhp=200$ GeV. The short-dashed and solid curve show
the predictions in the SM and model III using
$F_0^{B\pi}(0)=0.25$. The dots band corresponds to the (averaged)
data with $2\sigma$ errors.}
\label{fig:fig2}
\end{minipage}
\end{figure}

\newpage
\begin{figure}[t] 
\vspace{-60pt}
\begin{minipage}[t]{0.95\textwidth}
\centerline{\epsfxsize=\textwidth \epsffile{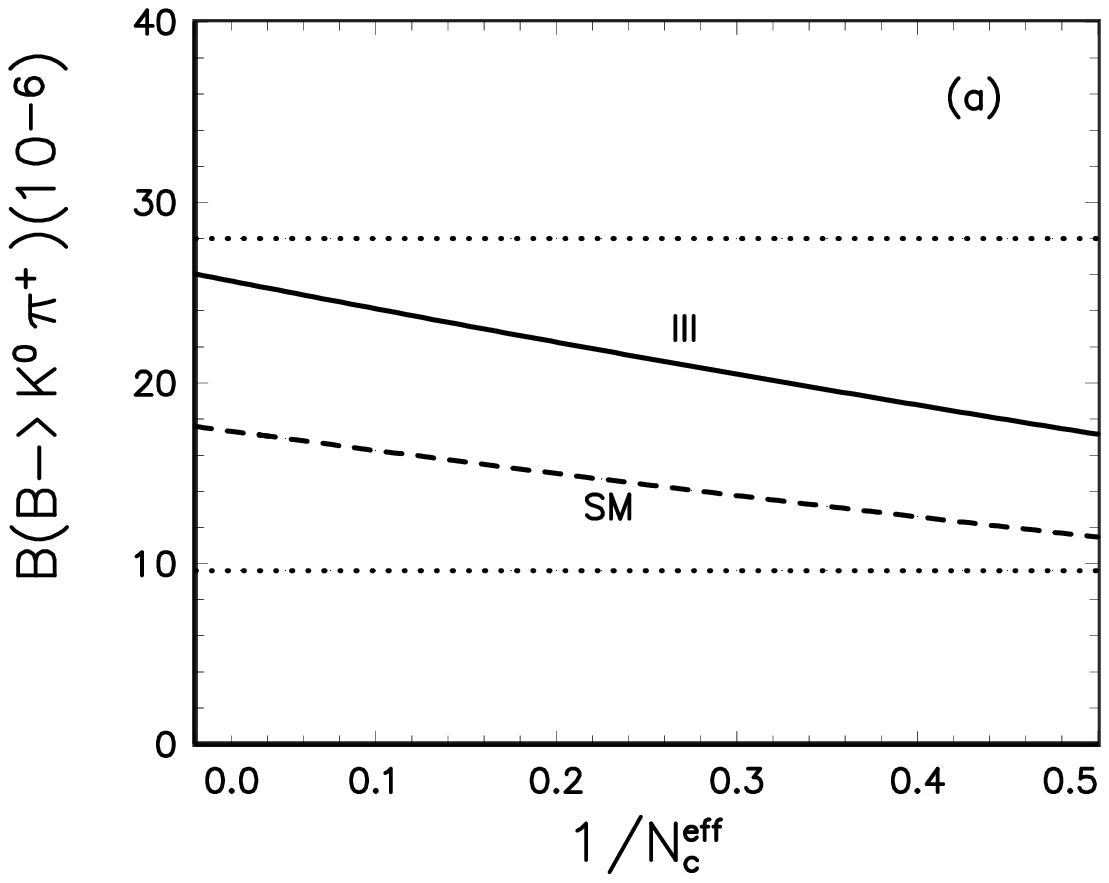}}
\vspace{-60pt}
\centerline{\epsfxsize=\textwidth \epsffile{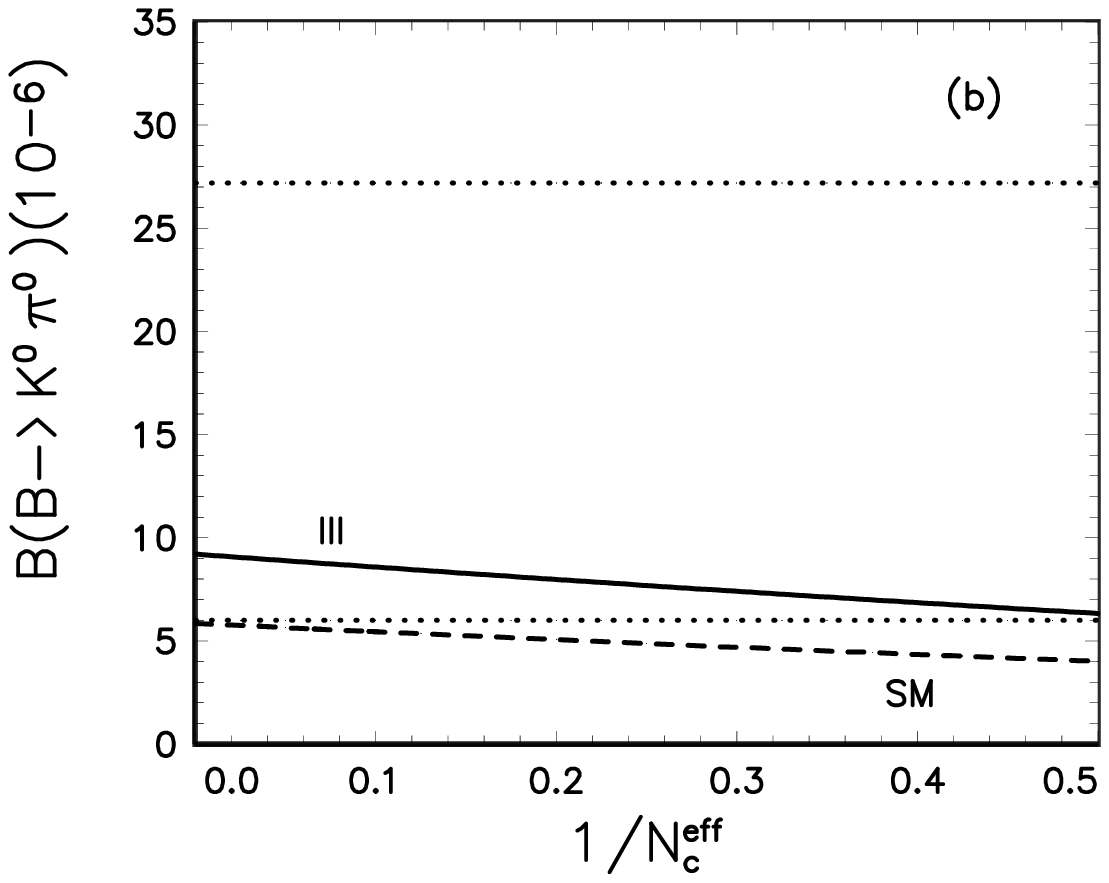}}
\caption{Same as \fig{fig:fig2} but for  $B \to K^0 \pi^+$ (3a) and
$K^0 \pi^0$ (3b) decay modes. }
\label{fig:fig3}
\end{minipage}
\end{figure}

\newpage
\begin{figure}[t]
\vspace{-60pt}
\begin{minipage}[t]{0.95\textwidth}
\centerline{\epsfxsize=\textwidth \epsffile{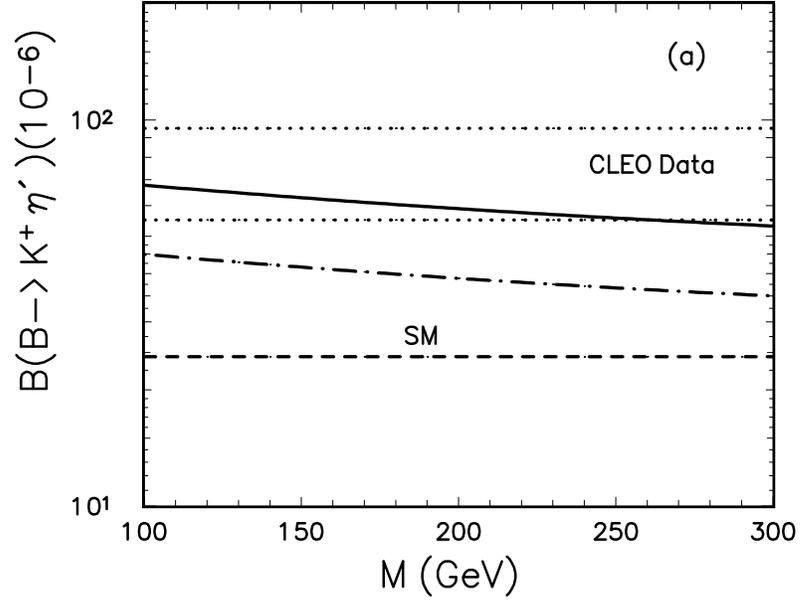}}
\vspace{-60pt}
\centerline{\epsfxsize=\textwidth \epsffile{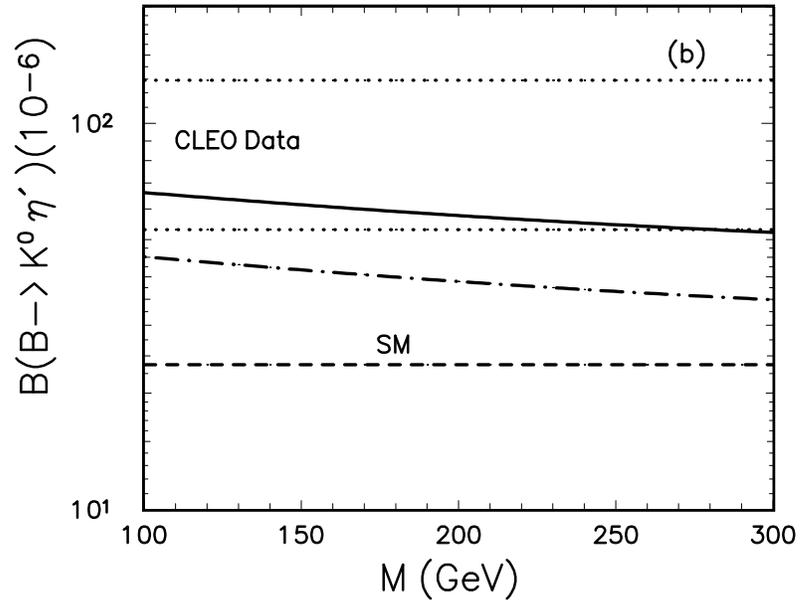}}
\caption{Same as \fig{fig:fig1} but for $F_0^{B\pi}(0)=0.25$.}
\label{fig:fig4}
\end{minipage}
\end{figure}

\newpage
\begin{figure}[t] 
\vspace{-60pt}
\begin{minipage}[t]{0.95\textwidth}
\centerline{\epsfxsize=\textwidth \epsffile{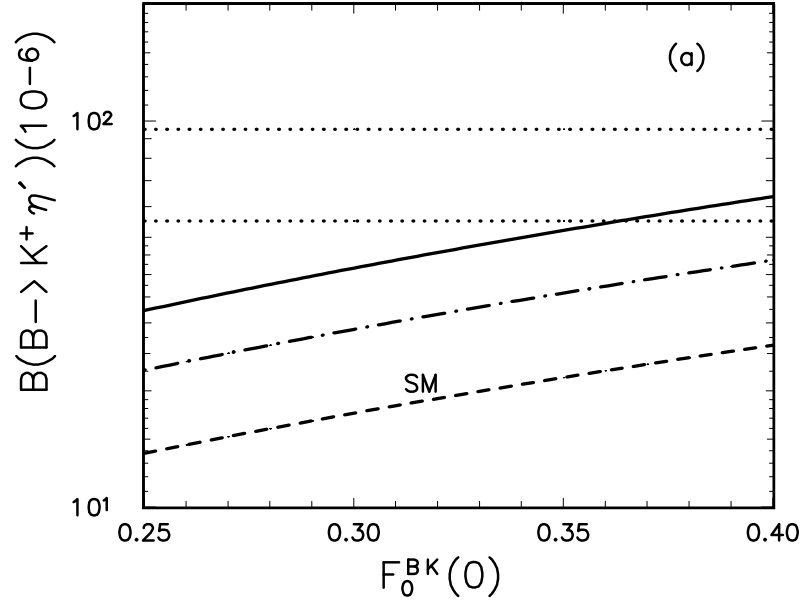}}
\vspace{-60pt}
\centerline{\epsfxsize=\textwidth \epsffile{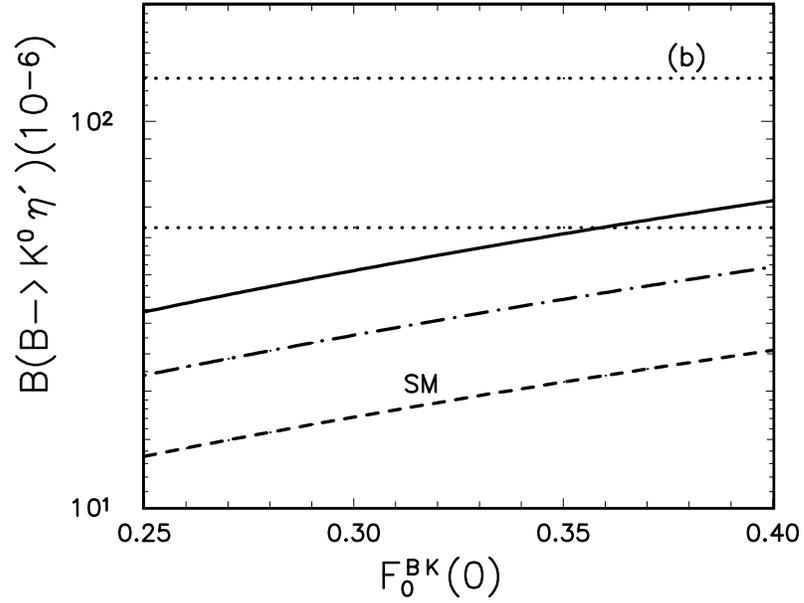}}
\caption{Plots of branching ratios of decays $B^+ \to K^+ \etap$ (5a) and
$B^0 \to K^0 \etap$ (5b) versus form factor $F_0^{BK}(0)$ in the SM and model III.
The short-dashed line shows the SM predictions with $N_c^{eff}=3$.
The dot-dashed and solid curve refers to the branching ratios in the model III
for $N_c^{eff}=3$ and $\infty$, respectively.
The dots band corresponds to the (averaged) data with $2\sigma$ errors. }
\label{fig:fig5}
\end{minipage}
\end{figure}

\newpage
\begin{figure}[t] 
\vspace{-60pt}
\begin{minipage}[t]{0.95\textwidth}
\centerline{\epsfxsize=\textwidth \epsffile{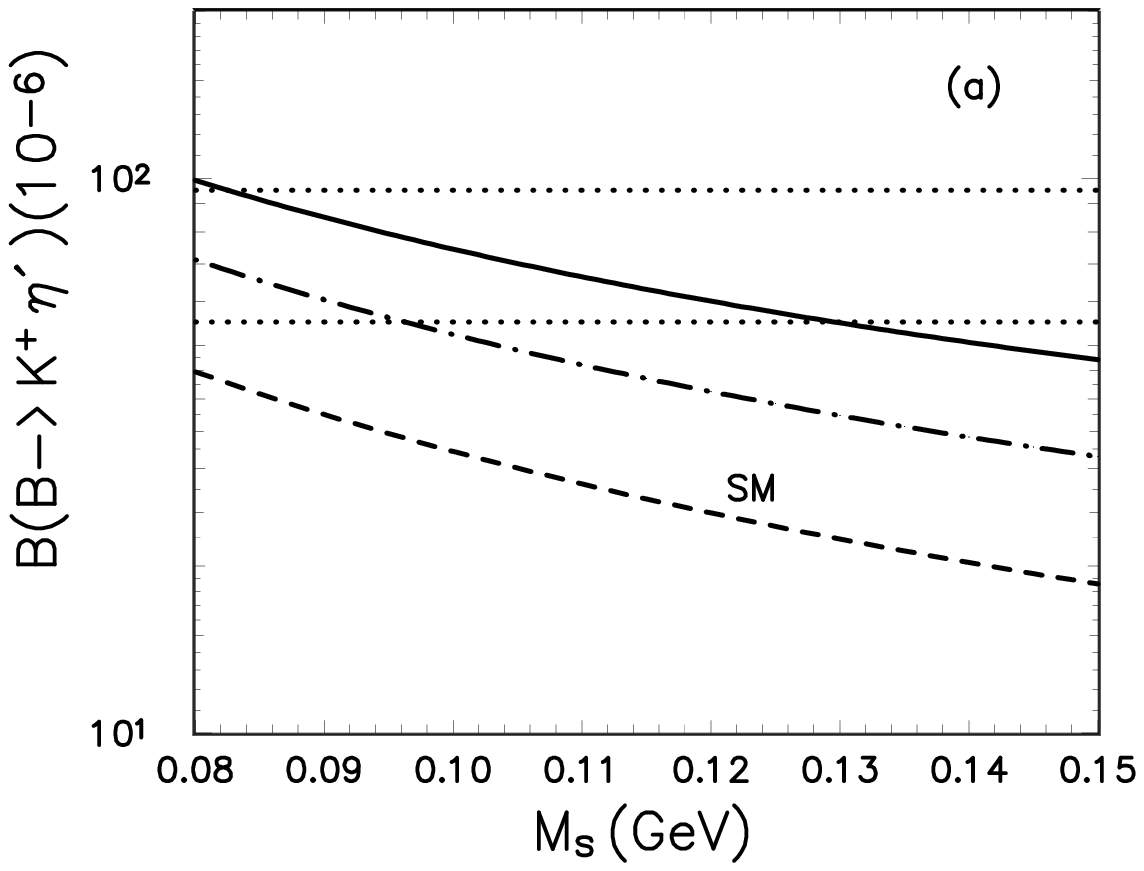}}
\vspace{-60pt}
\centerline{\epsfxsize=\textwidth \epsffile{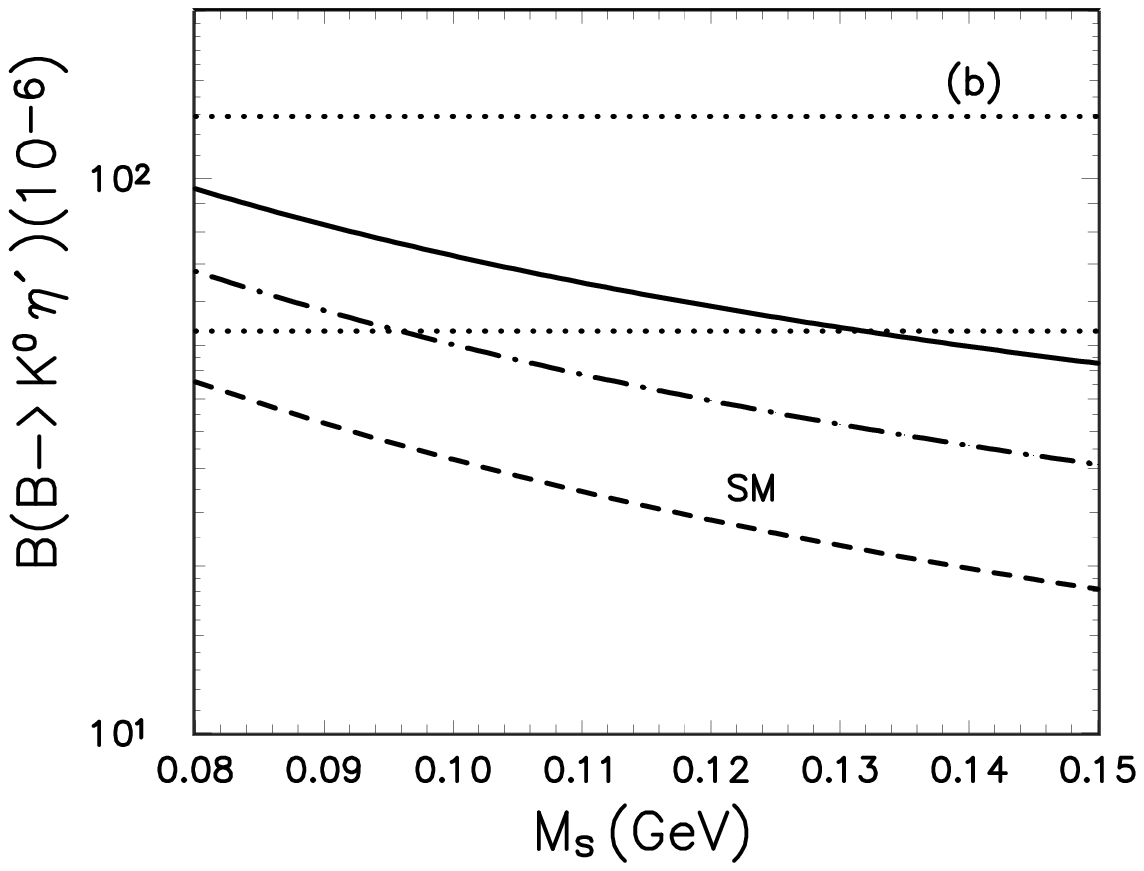}}
\caption{Plots of branching ratios of decays $B^+ \to K^+ \etap$ (6a) and
$B^0 \to K^0 \etap$ (6b) versus mass $m_s$ in the SM and model III
with $F_0^{B\pi}=0.25$, $F_0^{BK}=0.33$ and $\mhp=200$ GeV.
The short-dashed line shows the SM predictions with $N_c^{eff}=3$.
The dot-dashed and solid curve refers to the branching ratios in the model III
for $N_c^{eff}=3$ and $\infty$, respectively.
The dots band corresponds to the (averaged) data with $2\sigma$ errors. }
\label{fig:fig6}
\end{minipage}
\end{figure}

\end{document}